\documentclass[12pt]{article}
\usepackage{amscd,amssymb,amsmath,latexsym,enumerate}
\usepackage[mathscr]{euscript}
\usepackage{epsfig}
\usepackage{textcomp}
\usepackage{amsmath}
\usepackage{amssymb}
\usepackage{verbatim}

\title{The density of surface states as the total time delay}
\vspace{.5cm}

\author{Hermann Schulz-Baldes
\\
{\small Department Mathematik, Universit\"at Erlangen-N\"urnberg, D-91058
Erlangen, Germany} 
\\
{\small Email: schuba@mi.uni-erlangen.de} 
\vspace{.2cm}
}

\date{ }

\newtheorem{theo}{Theorem}
\newtheorem{defini}{Definition}
\newtheorem{proposi}{Proposition}
\newtheorem{lemma}{Lemma}
\newtheorem{coro}{Corollary}

\newcommand{\CM}{{\mathbb C}}

\newcommand{\NM}{{\mathbb N}}

\newcommand{\PM}{{\mathbb P}}
\newcommand{\RM}{{\mathbb R}}

\newcommand{\TM}{{\mathbb T}}

\newcommand{\ZM}{{\mathbb Z}}

\newcommand{\Aa}{{\cal A}}                      

\newcommand{\Ff}{{\cal F}}                      
\newcommand{\Ee}{{\cal E}}                      
\newcommand{\Uu}{{\cal U}}
\newcommand{\Oo}{{\cal O}}
\newcommand{\Rr}{{\cal R}}
\newcommand{\Ss}{{\cal S}}
\newcommand{\Dd}{{\cal D}}

\newcommand{\Tt}{{\cal T}}

\newcommand{\Jj}{{\mathcal J}}
\newcommand{\Kk}{{\mathcal K}}

\newcommand{\EE}{{\bf E}}
\newcommand{\Psur}{P_{\mbox{\rm\tiny sur}}}

\newcommand{\Ran}{\mbox{\rm Ran}}
\newcommand{\one}{{\bf 1}}
\newcommand{\Tr}{\mbox{\rm Tr}}
\newcommand{\TR}{\mbox{\rm Tr}}
\newcommand{\Ker}{\mbox{\rm Ker}}

\def\XXint#1#2#3{{\setbox0=\hbox{$#1{#2#3}{\int}$}
     \vcenter{\hbox{$#2#3$}}\kern-.5\wd0}}

\begin{document}

\maketitle

\begin{abstract} 
For a scattering problem of tight-binding Bloch electrons by a weak random surface potential,  a generalized Levinson theorem  is put forward showing the equality of the total density of surface states and the density of the total time delay. The proof uses explicit formulas for the wave operators in the new rescaled energy and interaction (REI) representation, as well as an index theorem for adequate associated operator algebras.
\\
\\
MSC 2010: 81U99, 47A40, 19K56 \hspace{1.5cm} Keywords: surface scattering, Levinson theorem
\end{abstract}




\vspace{.5cm}

\section{Main result and short discussion}
\label{sec-intro}

Let $H_0$ be a translation invariant finite distance hopping operator on $\ell^2(\ZM^{d})$ with only one energy band $[E_-,E_+]\subset\RM$ and $V$ a bounded and finite range surface perturbation supported on a subspace $\Lambda=\ZM^{d_1}\times\{0\}$ where $0$ denotes the zero vector $\ZM^{d_2}$ with $d_2={d}-d_1$. Both $d_1$ and $d_2$ are supposed to be positive. The perturbed Hamiltonian is $H=H_0+V$. It is well-known (Rayleigh, Tamm, Shockley and many others) that there are surface states for such Hamiltonians and there are many papers analyzing their spectral properties and the surface density of states, {\it e.g.} \cite{EKSS,JMP,JM,Cha,KS,KK}. The scattering problem for the pair $(H,H_0)$ was studied by Jaksic and Last \cite{JL1,JL2} who showed that the wave operators exist (this was proved independently by Chahrour and Sahbani \cite{CS}) and have common range so that the scattering operator is well-defined. The orthogonal complement of the range of the wave operators is then the subspace of surface states which can also be characterized as those states which do not diffuse  away from the boundary. These results from \cite{JL1,JL2} are described below. Focus will here be on a random family $(V_\omega)_{\omega\in\Omega}$ of surface perturbations satisfying a standard covariance property along the support $\Lambda$ of $V_\omega$ (see Section~\ref{sec-algebras}). Here $\Omega$ is a compact probability space equipped with a $\ZM^{d_1}$ action and an invariant and ergodic probability measure.  Then the Hamiltonians $H_\omega=H_0+V_\omega$ also form such a covariant family $H=(H_\omega)_{\omega\in\Omega}$. Typical examples are periodic, quasiperiodic and random surface potentials. For technical reasons explained below, the main result contains the unphysical hypothesis $d_2\geq 3$. Further below in the introduction is a discussion of what should hold without it.

\begin{theo}
\label{theo-LevinsonGen}
Suppose $d_2\geq 3$ and $\|V_\omega\|\leq C_0$ with a constant $C_0>0$ depending on $H_0$ and determined below. Consider the scattering problems $(H_\omega,H_0)$ and let $S_\omega$ be the associated scattering operator and  $P_{\mbox{\rm\tiny sur},\omega}$ the spectral projection of $H_\omega$ onto the surface states. Then $S=(S_\omega)_{\omega\in\Omega}$ and $\Psur=(P_{\mbox{\rm\tiny sur},\omega})_{\omega\in\Omega}$ are $\ZM^{d_1}$-covariant operator families on $\ell^2(\ZM^{d_1})$ and
\begin{equation}
\label{eq-main}
\Tt_1\,\Tr_2(\Psur)\;=\;
-\,\frac{1}{2\pi\imath}\int^{E_+}_{E_-} \!\!dE\;
\Tt_1\bigl((\overset{\;\circ}{S}_E)^*\,\partial_E\,\overset{\;\circ}{S}_E\bigr)
\;,
\end{equation}
where $\Tt_1\,\Tr_2$ is the trace per unit volume $\Tt_1$ along $\ZM^{d_1}$ combined with the usual trace $\Tr_2$ in the directions $\ZM^{d_2}$  transverse to the hypersurface, and $\overset{\;\circ}{S}_E=(\overset{\;\circ}{S}_{E,\omega})_{\omega\in\Omega}$ is the on-shell scattering matrix in the energy and interaction representation which is a $\ZM^{d_1}$-covariant family of unitary operators on $\ell^2(\ZM^{d_1})$ constructed below.
\end{theo}

The l.h.s. of the equality \eqref{eq-main} is the total density of surface states, and the r.h.s. is the total time delay density given as the non-commutative (non-integer) winding number of the path $E\mapsto \overset{\;\circ}{S}_E$ of unitaries in the crossed product C$^*$-algebra $C(\Omega)\rtimes \ZM^{d_1}$ of $\ZM^{d_1}$-covariant operators on $\ell^2(\ZM^{d_1})$. Hence the formula \eqref{eq-main} generalizes the well-known Levinson theorem connecting the number of bound states of a short range scattering problem to the total scattering phase. As pointed out in \cite{KR,KR2}, the Levinson theorem and hence also the identity \eqref{eq-main} can be seen to result from an index theorem connecting two index pairing and it is hence topologically very robust. Indeed, on the l.h.s. of \eqref{eq-main} one has the pairing of a $0$-cocylce $\Tt_1\,\Tr_2$ with a projection $\Psur$ specifying a $K_0$-class of the C$^*$-algebra $C(\Omega)\rtimes \ZM^{d_1}$, and on the r.h.s. of  \eqref{eq-main} is the pairing of a $1$-cocycle with a unitary in the C$^*$-algebra $C_0((E_-,E_+))\otimes C(\Omega)\rtimes \ZM^{d_1}$ which can also be interpreted as a spectral flow in a type II$_1$ von Neumann algebra. The two algebras and hence the two pairings are connected by an exact sequence, see Sections~\ref{sec-finitesupport} and \ref{sec-algebras}.

\vspace{.2cm}

Further results of the paper are formulas for the wave operators similar to those in \cite{KR,KR2,BS,RT}, as well as for the projection $\Psur$ and the scattering operator. They are given in the new rescaled energy and interaction (REI) representation which carries its name because the energy interval $[E_-,E_+]$ is rescaled to $\RM$ and the fixed energy fibers in this representation are the Hilbert space $\ell^2(\ZM^{d_1})$ associated to the support of the perturbation. On first sight, the REI representation may resemble the Schur complement formula used in \cite{JL1}, but it is in fact quite distinct. 


\vspace{.2cm}

Rather straight-forward generalizations (discussed briefly in Section~\ref{sec-periodic}) allow the perturbation $V$ to lie on an arbitrary hypersurface which is not a coordinate plane in $\ZM^d$ such as $\ZM^{d_1}\times\{0\}$. This is relevant for the $2$-magnon problem and its variations \cite{GS}. Furthermore, the techniques still transpose to the case where the perturbation $V$ has its support on a finite distance of the hypersurface, but this is not further developed here. The hypothesis that $d_2\geq 3$ is imposed (as in \cite{BS}) because the van Hove singularities of translation invariant operators are milder in higher dimension. In particular, the density of states of such an operator is continuous in dimension larger than or equal to three. The cases $d_2=1,2$ of physical interest can in principle also be dealt with by the formalism and the techniques of this paper, but further analysis of the Green function of $H_0$ restricted to the hypersurface is needed, that is Corollary~\ref{coro-freeGreen} has to be circumvented.

\vspace{.2cm}

One of the remarkable features of \eqref{eq-main} is that the surface spectrum may have non-trivial intersection with the spectrum $[E_-,E_+]$ of the free operator $H_0$. As an example for superposed absolutely continuous surface spectrum, the case of a constant surface perturbation is discussed in Section~\ref{sec-periodic}.  For this example, the equality \eqref{eq-main} is also shown to hold without the weak coupling hypothesis $\|V_\omega\|\leq C_0$. In the general case of a covariant surface perturbation, this hypothesis is needed as a technical input for the calculation of the wave operators. The difficulties arising without this hypothesis are discussed at the end of  Section~\ref{sec-wavecalc}. In the case of a perturbation with finite support these difficulties were overcome in \cite{BS} and this allowed to deal with embedded eigenvalues and half-bound states. 

\vspace{.2cm}

Let us hint at another question left open in this paper. As in \cite{JL2}, the projection on the surface states is given by $\Psur=\one-W_\pm W_\pm^*$ where $W_\pm$ are the wave operators and then $\Tt_1\TR_2(\Psur)$ the density of these surface states. On the other hand, in numerous prior works \cite{EKSS,Cha,KS} an adequate spectral shift function was used to define a surface state density (which then has no definite sign). The relation between the two notions is not clarified here. It is reasonable to expect a link via a so-called spectral property of the time delay ({\it e.g.} Section~4.7 of \cite{BS}). This would also be in line with \cite{KKN}.

\vspace{.2cm}

The author expects that \eqref{eq-main} also holds in a strong coupling regime and for $d_2=1,2$, possibly with a corrective term stemming from half-bound states. This would then establish that the link between total surface state and scattering phase densities holds irrespective of the spectral nature of the surface states. Indeed the above weak  coupling hypothesis implies that $H$ has no singular spectrum at all, see \cite{JL1} and Section~\ref{sec-spectral}. On the other, for half-space models with random surface potentials in $d=2$ \cite{JM} as well as for $d\geq 3$ and an either weak or strong random surface potential (but not an intermediate one) \cite{JM2}, the surface spectrum is purepoint outside of the spectrum of $H_0$. However, these results for half-space models do not transpose directly to the models considered here.

\vspace{.2cm}

This work is an extension of the prior joint work with Jean Bellissard \cite{BS} which treated the scattering problem for lattice operators $H_0$ perturbed by perturbations $V$ with finite support. The techniques of this prior work are heavily used here and the reader may be forced to go back to it for proofs of some technical facts.  However, the present work contains one crucial technical addendum to \cite{BS}, namely the REI representation of the main operators of scattering theory. Implicitly, this was contained in \cite{BS}, but here it is formalized. It is only in the REI representation that the covariance properties of the perturbation $V$ can be used  for the scattering problem. It also allows to use a more simple exact sequence of operator algebras for the proof of Levinson's theorem in the case of a finitely supported perturbation. This is explained in Section~\ref{sec-finitesupport} where also an obvious mistake in the statement of Levinson's theorem made in \cite{BS} is corrected. Further minor corrections to \cite{BS} will also be mentioned.

\vspace{.2cm}

\noindent {\bf Acknowledgements:} Apart from the collaboration with Jean Bellissard, the author profited from several discussions with Magnus Goffeng and Miguel Ballesteros. He also thanks the Mittag-Leffler Institute, the Instituto de Matem\'aticas de la UNAM, Unidad Cuernavaca, and the DFG for support.

\section{Case of a constant surface perturbation}
\label{sec-periodic}

As a motivation for the sequel and also in order to introduce some notations, let us provide the proof of \eqref{eq-main} for the case of a constant surface potential as an example. It will be argued below that this covers models related to the $2$-magnon problem \cite{GS} in the context of which the surface states are also called bound states. Let us begin with a detailed description of the translation invariant operator $H_0$ on $\ell^2(\ZM^{d})$. It is supposed to be of the form
\begin{equation}
\label{eq-Ham0}
H_0\;=\;\sum_{{n}\in\ZM^{d}}\Ee_{n}\,U_{n}\;,
\end{equation}
where $U_{n}$ denotes the translation operators by ${n}\in\ZM^d$ on $\ell^2(\ZM^{d})$ and $\Ee_{n}=\overline{\Ee_{ -n}}\in\CM$ are coefficients with exponential decay in $n$ such that 
$$
\Ee({z})
\;=\;
\sum_{{n}\in\ZM^{d}}
\Ee_{n}\;{z}^{n}
\;,
$$
is analytic on a neighborhood of the torus $\TM^{d}\subset\CM^{d}$.  It is supposed that the $H_0$ acts non-trivially in all directions of $\ZM^d$. Due to the symmetry $\Ee_{n}=\overline{\Ee_{ -n}}$, the function $\Ee$ is real on $\TM^d$. Abusing notations, we also simply write $\Ee({k})=\Ee(e^{\imath {k}})$. Then the discrete Fourier transform 
$$
{\bf \Ff}:\ell^2(\ZM^{d})\to L^2(\TM^{d})
\;,
\qquad
({\bf \Ff}\phi)({k})\;=\;
(2\pi)^{-\frac{{d}}{2}}\;
\sum_{{n}\in\ZM^{d}} e^{\imath {k} \cdot n}\,\phi_{{n}}\;,
$$
diagonalizes $H_0$:
$$
({\bf \Ff}\,H_0\,{\bf \Ff}^*\phi)_{k}
\;=\;
\Ee({k})\;\phi_{k}
\;,
\qquad
\phi\in L^2(\TM^{d})
\;.
$$
Here ${k} \cdot n=\sum_{j=1}^d k(j)n(j)$ denotes the euclidean scalar product, expressed in terms of the components $k(j)$ and $n(j)$ of $k$ and $n$. The standard example is the discrete Laplacian for which $\Ee({k})=2\sum_{j=1}^{d}\cos(k(j))$.  Let us set $E_-=\min(\Ee)$ and $E_+=\max(\Ee)$ and suppose that these are the only local extrema of $\Ee$. Also the partial Fourier transform will be used:
$$
\Ff_1:\ell^2(\ZM^{d})\to L^2(\TM^{d_1})\otimes\ell^2(\ZM^{d_2})
\;,
\qquad
(\Ff_1\phi)_{n_2}(k_1)\;=\;
(2\pi)^{-\frac{d_1}{2}}\;
\sum_{n_1\in\ZM^{d_1}} e^{\imath k_1\cdot n_1}\,\phi_{(n_1,n_2)}\;.
$$
Then
\begin{equation}
\label{eq-partialfib}
\Ff_1\,H_0\,\Ff_1^*
\;=\;
\int^\oplus_{\TM^{d_1}} dk_1\;H_{0}(k_1)
\;,
\end{equation}
where $H_{0}(k_1)$ is an operator on $\ell^2(\ZM^{d_2})$ given by
$$
H_{0}(k_1)
\;=\;
\sum_{n_2\in\ZM^{d_2}}\Ee_{n_2}(k_1)\,U_{n_2}
\;,
$$
with translation operators $U_{n_2}$ on $\ell^2(\ZM^{d_2})$ and 
$$
\Ee_{n_2}(k_1)
\;=\;
\sum_{n_1\in\ZM^{d_1}}
\Ee_{(n_1,n_2)}\;e^{\imath k_1\cdot n_1}
\;.
$$

\vspace{.2cm}

Now $H_0$ will be perturbed by a bounded operator $V$ on $\ell^2(\ZM^{d})$ supported on the subset  $\Lambda\subset\ZM^d$. Associated to $\Lambda$ is in a natural way the subspace of $\ell^2(\ZM^{d})$ of the states supported by $\Lambda$. Let $\Pi:\ell^2(\ZM^{d})\to \ell^2(\Lambda)$ be the associated partial isometry, namely $\Pi^*\Pi$ is the projection in $\ell^2(\ZM^{d})$ onto the subspace and $\Pi\,\Pi^*=\one_{\ell^2(\Lambda)}$. Then the perturbation satisfies $V=\Pi^*\Pi V\Pi^*\Pi$ which means nothing but that $V$ is supported by $\Lambda$.  In our previous work \cite{BS} is was supposed that $\Lambda$ is finite. Even though many results below also hold for finite $\Lambda$, the focus here is mainly on $\Lambda=\ZM^{d_1}\times\{0\}$ where $0$ denotes the zero vector $\ZM^{d_2}$ with $d_2={d}-d_1$. Both $d_1$ and $d_2$ are supposed to be positive. Then $V$ is called a surface perturbation, and in case it is diagonal in position space, $V$ is called a surface potential and in this section only constant surface potentials are considered. 

\vspace{.2cm}

The case of a constant surface potential is of interest for the so-called $2$-magnon problem \cite{GS}. Here $d_1=d_2$ so that ${d}=2d_1$. The potential is then rather on the diagonal $\{(n,n)\,|\,n\in\ZM^{d_1}\}\subset \ZM^{d}$, but by the bijection $\varphi:\ZM^{d}\to\ZM^{d}$ given by $\varphi(n_1,n_2)=(n_1,n_2-n_1)$ this diagonal is mapped to the first component so that one is again in the case above. If $H_0$ is the discrete Laplacian in the setting before this transformation (as it is the case in the $2$-magnon problem), then after the transformation it is not the discrete Laplacian any more, but it is still translation invariant and of finite range, and thus of the form \eqref{eq-Ham0} given above.  

\vspace{.2cm}

Let now the constant surface potential be $V=\lambda\,\Pi^*\Pi$ on $\Lambda=\ZM^{d_1}\times\{0\}$. In this situation, both $H_0$ and $H=H_0+V$ are partially diagonalized by $\Ff_1$:
\begin{equation}
\label{eq-partialfibH}
\Ff_1\,H\,\Ff_1^*
\;=\;
\int^\oplus_{\TM^{d_1}} dk_1\;H(k_1)
\;,
\qquad
H(k_1)\;=\;H_0(k_1)\,+\,\lambda\,|0\rangle\langle 0|
\;,
\end{equation}
where $|0\rangle$ is the state at the origin in $\ell^2(\ZM^{d_2})$. Thus one has a scattering problem for  the pair $(H_0(k_1),H(k_1))$ for each fixed $k_1\in\TM^{d_1}$, which hence respects the fibration. The bound states of $H(k_1)$ constitute the surface states and for almost all $k_1$ there are no half-bound states. This scattering problem can be analyzed by the (elementary) techniques of \cite[Section~3.9]{BS}. As an example, let us consider the discrete Laplacian $H_0$ in dimension $d=3$, and let $d_1=1$ and $d_2=2$. Then $H(k_1)=H_0(k_1)+\lambda\,|0\rangle\langle 0|+2\,\cos(k_1)$ where $H_0(k_1)$ is a $2$-dimensional discrete Laplacian. Its resolvent $G_0(k_1,E-\imath\,0)=\langle 0|(H_0(k_1)-E+\imath\,0)^{-1}|0\rangle$ then has a logarithmically divergent real part as $E$ approaches the band edges from outside. Hence ({\it e.g.} by the argument of Section~\ref{sec-spectral}) pending on the sign of $\lambda$ there are bound states for $H(k_1)$ above or below the energy band of the free operator $H_0(k_1)$. As $k_1$ varies in $[-\pi,\pi)$ this  leads to a band of surface states which energetically have one part lying outside of the band of $H_0$ and another part inside (over) the band of $H_0$. Let us point out that approximating $V$ by $\lambda\sum_{|n_1|\leq N_1}|n_1\rangle\langle n_1|$ does not lead to bound states for any $N_1\in\NM$ and $\lambda$ sufficiently small (by Section~\ref{sec-spectral}) so that the spectra do not converge in the limit $N_1\to\infty$.

\vspace{.2cm}

Under the supplementary hypothesis that each $H_0(k_1)$ has only two local extrema, the hypothesis of \cite{BS} are satisfied so that Levinson's theorem holds (without half-bound states). This is rederived in Section~\ref{sec-finitesupport} below. It shows that the number $N(k_1)\in\{0,1\}$ of bound states of $H(k_1)$ is equal to the total scattering phase
$$
N(k_1)
\;=\;
-\,\frac{1}{2\pi\imath}\;
\int^{E_+(k_1)}_{E_-(k_1)}\!\!dE\;
\Tr\left(\overset{\;\circ}{S}_E(k_1)^*\partial_E \overset{\;\circ}{S}_E(k_1)\right)
\;=\;
-\,\frac{1}{2\pi\imath}\;
\int\!db\;
\Tr\left(\overset{\;\circ}{S}_b(k_1)^*\overset{\;\circ}{S}_b(k_1)\right)
\;,
$$
where $\overset{\;\circ}{S}_E(k_1)$ and $\overset{\;\circ}{S}_b(k_1)$ are the scattering matrices in the EI and REI representations as constructed below. Alternatively, the EF and REF representations of \cite{BS} can be used to deduce these formulas. These operators act on the one-dimensional Hilbert space span$(|0\rangle)\cong\CM$ so that the trace can be dropped.   Now let us integrate over $k_1$:
\begin{equation}
\label{eq-Levinsonfibered}
\int_{\TM^{d_1}}\frac{dk_1}{(2\pi)^{d_1}}\;N(k_1)
\;=\;
-\,
\frac{1}{2\pi\imath}\;
\int_{\TM^{d_1}}\frac{dk_1}{(2\pi)^{d_1}}\;
\int\!db\;
\Tr\left(\overset{\;\circ}{S}_b(k_1)^*\partial_b \overset{\;\circ}{S}_b(k_1)\right)
\;.
\end{equation}
The l.h.s. of this formula can be rewritten in a more conceptual and compact way using the following tracial state defined for covariant operators $O$ on $\ell^2(\ZM^{d})$:
\begin{equation}
\label{eq-Ttrace}
\Tt_1\,\Tr_2(O)
\;=\;
\lim_{N\to\infty}
\;
\frac{1}{(2N+1)^{d_1}}
\;
\Tr\left(
\chi_N\,O\,\chi_N
\right)
\;,
\end{equation}
where $\chi_N$ is the indicator function on the box $[-N,N]^{d}$ in dimension ${d}$. Note that $\Tt_1\,\Tr_2(\one)=\infty$, but $\Tt_1\,\Tr_2(\Pi^*\Pi)=1$. The state $\Tt_1\,\Tr_2$ is the trace per unit volume along $\Lambda$, but the usual trace in the perpendicular direction. Its definition extends to covariant operators. For an operator $O$ that is translation invariant along $\Lambda$ (such as $H_0$ and, for the situation in this section, also $H$), one has
$$
\Tt_1\,\Tr_2(O)
\;=\;
\int_{\TM^{d_1}}\frac{dk_1}{(2\pi)^{d_1}}\;
\Tr(O(k_1))
\;,
\qquad
\Ff_1\,O\,\Ff_1^*\;=\;
\int^\oplus_{\TM^{d_1}} dk_1\;O(k_1)
\;.
$$
As in the present situation, the projection $\Psur$ on the surface states is of this fibered form, one concludes that the formula \eqref{eq-Levinsonfibered} can be rewritten as 
$$
\Tt_1\,\Tr_2(\Psur)\;=\;
-\,
\frac{1}{2\pi\imath}\;
\int\!db\;
\int_{\TM^{d_1}}\frac{dk_1}{(2\pi)^{d_1}}\;
\Tr\left(\overset{\;\circ}{S}_b(k_1)^*\partial_b \overset{\;\circ}{S}_b(k_1)\right)
\;.
$$
This formula is the same as in Theorem~\ref{theo-LevinsonGen}. The main aim of the paper is to prove this formula also for covariant surface potentials.

\section{Analysis of the unperturbed operator}
\label{sec-unperturbed}

\subsection{Dilation operator and REF representation}
\label{sec-dilop}

This section merely reviews results and notations from \cite{BS}. Let $E_-$ and $E_+$ be the boundaries of the spectrum of $H_0$ and set
$$
F(E)\;=\;2\;
\frac{(E-E_-)(E_+-E)}{E_+-E_-}\,,
\qquad
f(E)\;=\;
 \int^E_{E_r}\frac{de}{F(e)}
\;=\;
  \frac{1}{2}\,
   \ln\left(\frac{E-E_-}{E_+-E}\right) 
 \,,
$$
where $E_r=\frac{1}{2}(E_+-E_-)$ is some reference energy.  Then a vector field $\widehat{X}$ on $\TM^d$ is defined by
$$ 
\widehat{X}(k) \;=\;
F\bigl(\Ee(k)\bigr)\;
   \frac{\nabla \Ee(k)}{|\nabla\Ee(k)|^2}\;,\qquad k\in\TM^d\;.
$$
Let $X_j=\Ff^*\widehat{X}_j\Ff$ be the operator on $\ell^2(\ZM^d)$ associated with the $j$th component $\widehat{X}_j$ of $\widehat{X}$. Also let $Q=(Q_1,\ldots,Q_d)$ be the position operator defined by $Q_j\,\phi(n) =n_j\,\phi(n)$, for $n\in \ZM^d$ and $\phi$ decreasing sufficiently fast. Then set
\begin{equation}
\label{eq-dilation}
A\;=\;\frac{1}{2}\,
 \sum_{j=1}^d
  \left( X_j\,Q_j+Q_j\,X_j\right)\,.
\end{equation}
This defines a self-adjoint operator satisfying
$$
\imath [A,H_0]\;=\;F(H_0)\;.
$$
Furthermore, the Fourier transform of the associated strongly continuous one-parameter unitary group is explicitly given by
$$
(e^{\imath b \widehat{A}}\,\phi)(k)\;=\;
\det(\theta'_b(k))^{\frac{1}{2}}
\;\phi(\theta_b(k))\;=\;
 \exp\left(
    \frac{1}{2}
     \int_0^b du\;
      \mbox{div}(\widehat{X})(\theta_u(k))
     \right)\;
      \phi(\theta_b(k))\,,
$$
\noindent where $\theta_b:\TM^d\to\TM^d$ denotes the flow of the vector field $\widehat{X}$. Now associated to the reference energy  $E_r$ let us introduce the reference Fermi surface $\Sigma=\Ee^{-1}(E_r)$ with Riemannian volume $\nu$ obtained by restricting the Lebesgue measure to $\Sigma$. The coarea formula leads to the following change of variables (for adequate functions $\phi$):
$$
\int_{\TM^d} dk\;\phi(k)  
\;=\;
\int_\RM db
 \int_{\Sigma}\nu(d\sigma)\;
  \exp\left(
       \int_0^b du \;\mbox{div}(\widehat{X})(\theta_u(\sigma))
        \right)\;
   \Big|\widehat{X}(\sigma)\Big|\;
    \phi\left(\theta_b(\sigma)\right)\,.
$$
Therefore a unitary $\Uu:L^2(\TM^d)\to L^2(\RM)\otimes L^2(\Sigma,\nu)$ is (densely) defined by
$$
(\Uu \phi)_{b}(\sigma)\;=\;
 d_{b}(\sigma)\;
 \phi(\theta_{b}(\sigma))\,,
$$
where the following notation has been used:
$$
d_{b}(\sigma)
\;=\;
 \Big|\det(\theta'_b|_{T_\sigma\Sigma})\Big|^{\frac{1}{2}}\;
  \Big|\widehat{X}(\theta_b(\sigma))\Big|^{\frac{1}{2}}\;=\;
   \exp\left(
      \frac{1}{2}\;\int_0^{b} du \;\mbox{div}(\widehat{X})(\theta_u(\sigma))
       \right)\;
    \Big|\widehat{X}(\sigma)\Big|^{\frac{1}{2}}\,.
$$
The representation induced by $\Uu$ is called the rescaled energy and Fermi surface (REF) representation and $b=f(E)\in\RM$ is called the rescaled energy. Operators in this representation will be denoted by $\widetilde{O}=\Uu\,\widehat{O}\,\Uu^*$ where $\widehat{O}=\Ff\, O\,\Ff^*$. However, for sake of simplicity we will deviate from this notation in the case of the rescaled energy operator $B=\widetilde{B}$ and the dilation operator $A=\widetilde{A}=-\imath\,\partial_b$ in the REF representation.

\vspace{.2cm}

In \cite{BS} also the energy and Fermi surface (EF) representation was used. It is the REF representation, but with energy variable $E=f^{-1}(b)$ instead of $b$ so that EF represented operators act on the Hilbert space $L^2([E_-,E_+])\otimes L^2(\Sigma,\nu)$. Operators in the EF representation will have indices $E$ and $E'$ instead of $b$ and $b'$, but the tilde will be maintained (other than in \cite{BS}). Let us note that that $db=f'(E)dE$ and $\partial_b=f'(E)^{-1}\partial_E$. 

\subsection{Restricted free resolvent}
\label{sec-restricted}

The restricted free resolvent is defined by
\begin{equation}
\label{eq-freeresol}
G^\Lambda_0(z)
\;=\;
\Pi\,(z-H_0)^{-1}\,\Pi^*
\;,
\qquad
\Im m(z)\not = 0\;.
\end{equation}
(Let us note that, unfortunately, in \cite{BS} the role of $\Pi$ and $\Pi^*$ is erroneously exchanged at several places.) It is a bounded operator on $\ell^2(\Lambda)$ having the Herglotz property, so that it is invertible for $\Im m(z)\neq 0$. The following result is proved in  \cite{BS}.

\begin{proposi}
\label{prop-freeGreen} Let ${d}\geq 3$ and let $\Lambda$ be finite. The limits $G^\Lambda_0(E\pm\imath 0)=\lim_{\epsilon\downarrow 0} G^\Lambda_0(E\pm\imath \epsilon)$ exist.  Away from the critical values of $\Ee$, the map $E\in\RM\mapsto G^\Lambda_0(E\pm\imath 0)$ is real analytic. At the critical points it is H\"older continuous. Furthermore:

\begin{enumerate}[\rm (i)]

\item $\Im m\bigl(G^\Lambda_0(E\pm\imath 0)\bigr)$ vanishes on $(-\infty,E_-]\cup[E_+,\infty)$ and is positive semi-definite on $[E_-,E_+]$. Close to the band edges, one has
$$
\Im m\bigl(G^\Lambda_0(E-\imath 0)\bigr)
\;=\;
\Oo\bigl(|E-E_\pm|^{\frac{d}{2}-1}\bigr)
\;.
$$

\item The map $E\in\RM\mapsto \Re e\bigl(G^\Lambda_0(E)\bigr)$ is negative and decreasing on $(-\infty,E_-]$ and positive and decreasing on $[E_+,\infty)$. Furthermore, $G^\Lambda_0(\pm\infty)=0$.

\end{enumerate}

\end{proposi}

\begin{coro}
\label{coro-freeGreen} Let $d=d_1+d_2$ and $\Lambda=\ZM^{d_1}\times\{0\}$. Suppose ${d}_2\geq 3$. Then the weak limits $G^\Lambda_0(E\pm\imath 0)=\lim_{\epsilon\downarrow 0} G^\Lambda_0(E\pm\imath \epsilon)$ exist and are weakly H\"older continuous in $E\in\RM$. There exists a constant $C_0$ such that
\begin{equation}
\label{eq-GreenGlobBound}
\sup_{E\in\RM}\;
\left\|G^\Lambda_0(E)\right\|\;\leq\;(C_0)^{-1}\;.
\end{equation}
\end{coro}

\noindent {\bf Proof.} Due to \eqref{eq-partialfib} one has
\begin{equation}
\label{eq-GreenGlobFourier}
\Ff_1\,G^\Lambda_0(z)\,\Ff_1^*
\;=\;
\int^\oplus_{\TM^{d_1}} dk_1\;\langle 0|(z-H_{0}(k_1))^{-1}|0\rangle
\;,
\end{equation}
where $0\in\ZM^{d_2}$. For each $k_1$, the appearing matrix elements have limits $z=E\pm\imath 0$ by Proposition~\ref{prop-freeGreen} due to the hypothesis $d_2\geq 3$. Therefore a compactness argument in $k_1$ combined with Proposition~\ref{prop-freeGreen}  implies the bound \eqref{eq-GreenGlobBound}.
\hfill $\Box$

\vspace{.2cm}

It will be useful to characterize the kernel of $\Im m\bigl(G^\Lambda_0(E-\imath 0)\bigr)\geq 0$ for $E\in (E_-,E_+)$. It is a subspace of $\ell^2(\Lambda)$ and its orthogonal complement will be denoted by
\begin{equation}
\label{eq-Fbdef}
\Ff^\Lambda_b
\;=\;
\Ran\bigl(\,\Im m\;G^\Lambda_0(E-\imath 0)\bigr)
\;,
\qquad
b=f(E)
\;.
\end{equation}
Because $H_0$ is translation invariant and $\Pi$ is invariant under the subgroup $\ZM^{d_1}\subset\ZM^d$, the subspaces $\Ff^\Lambda_b$ is invariant under the action of $\ZM^{d_1}$. The orthogonal projection in $\ell^2(\ZM^{d_1})$ on $\Ff^\Lambda_b$ is denoted by $P_b^\Lambda $.  The following result parallels those in Section~2.7 of \cite{BS}, and shows that for many rescaled energies $b$, the dimension of $(\Ff^\Lambda_b)^\perp=\Ker\bigl(\,\Im m\;G^\Lambda_0(E-\imath 0)\bigr)$ is infinite (here the orthogonal complement is taken in $\ell^2(\Lambda)$). 

\begin{proposi}
\label{prop-Greenkernel} Suppose $\Lambda=\ZM^{d_1}\times\{0\}$. Let $E\in (E_-,E_+)$ be non-critical and denote the projection of the level surface ${\Sigma}_E=\{{k}\in\TM^{d}\,|\,\Ee({k})=E\}$ along the first component by
$$
\Sigma_{E,1}
\;=\;
\left\{k_1\in\TM^{d_1}\,|\,\mbox{\rm there exists }k_2\in\TM^{d_2}\;\mbox{\rm such that } (k_1,k_2)\in{\Sigma}_E
\right\}
\;.
$$
Then for $b=f(E)$
$$
(\Ff^\Lambda_b)^\perp
\;=\;
\Ker\;\Im m\bigl(G^\Lambda_0(E-\imath 0)\bigr)
\;=\;
\left\{
v\in\ell^2(\ZM^{d_1})\;\left|\;
\hat{v}(k_1)=0\;\mbox{\rm for almost all }k_1\in\Sigma_{E,1}
\right.
\right\}
\;,
$$
where $\hat{v}(k_1)\;=\;\sum_{n_1\in\ZM^{d_1}}v_{n_1}\,e^{\imath n_1\cdot k_1}$. Hence $\Ff_1 P^\Lambda_b\Ff_1^*=1-\chi_{\Sigma_{E,1}}$ in terms of the indicator function.  
\end{proposi}

\noindent  {\bf Proof:} Let us use the coarea formula and then the Plemelj-Privalov theorem (see \cite{BS} for details):
\begin{eqnarray*}
\langle v|\,\Im m\;G^\Lambda_0(E\pm\imath0)\,|v\rangle
& = &
\frac{1}{2\imath}\int^{E_+}_{E_-} de\;
\left(\frac{1}{E\pm\imath 0-e}-\frac{1}{E\mp\imath 0-e}\right)\;\int_{{\Sigma}_{e}}
\frac{\nu_e(d\sigma)}{(2\pi)^{d}}\;
\frac{|\hat{v}(\sigma)|^2}{|\nabla \Ee(\sigma)|}
\\
& = &
\mp \pi\;
\int_{{\Sigma}_{E}}
\frac{\nu_E(d\sigma)}{(2\pi)^{d}}\;
\frac{|\hat{v}(\sigma)|^2}{|\nabla \Ee(\sigma)|}
\;,
\end{eqnarray*}
where $\nu_E$ is the Riemannian measure ${\Sigma}_E$. Now, clearly the last integral only vanishes if $\hat{v}$ vanishes on the energy surface ${\Sigma}_E$. As $\hat{v}(\sigma)=\hat{v}(k_1)$ does not depend on the second component $k_2$ of $\sigma=(k_1,k_2)$, this proves the  statement. 
\hfill $\Box$

\subsection{REI representation}
\label{sec-localized}

This section is about the rescaled energy and interaction (REI) representation which is associated to $H_0$ and a subset $\Lambda\subset\ZM^d$ that is the support of the perturbation. It is {\em not } given by a unitary transformation of Hilbert space (such as $\Ff$ and $\Uu$ above), but rather by a partial isometry onto an adequate subspace. As it will turn out later on, the wave operator and other operators of scattering theory act non-trivially only on this subspace and therefore they will have an REI representation.

\vspace{.2cm}
 
Let us start with the REF representation of the localized state at site $m\in\ZM^d$ given by $\psi_m=\Uu\Ff\,|m\rangle$.  The states $(\psi_m)_{m\in\ZM^d}$ form an orthonormal basis in $L^2(\RM)\otimes L^2(\Sigma,\nu)$. More explicitly, they are given by
\begin{equation}
\label{eq-ONbasis}
\psi_{m,b}(\sigma)\;=\;
(2\pi)^{-\frac{d}{2}}\;
  d_b(\sigma)\;
   e^{\imath m\cdot \theta_b(\sigma)}\,,
\end{equation}
for almost all $\sigma\in\Sigma$. We now consider $\psi_{m,b}$ as a state in $L^2(\Sigma,\nu)$. These restricted localized states are not normalized, but their norm is independent of $m$. Then $(\psi_{m,b})_{m\in\ZM^d}$ is almost surely in $b$ a complete set in $L^2(\Sigma,\nu)$ because assuming the contrary readily leads to a contradiction. Furthermore
\begin{equation}
\label{eq-deltarel}
\sum_{m\in\ZM^d}\,|\psi_{m,b}\rangle\langle \psi_{m,b'}|
\;=\;
\one_{L^2(\Sigma,\nu)}\;\delta(b-b')
\;,
\end{equation}
if both sides are understood as integral kernels for operators on $L^2(\RM)$ with values in the bounded operators on $L^2(\Sigma,\nu)$. As by Lemma~2 of \cite{BS},
\begin{equation}
\label{eq-scalprodL2}
\langle\psi_{n,b}|\psi_{m,b}\rangle_{L^2(\Sigma,\nu)}\;=\;
   \frac{F(E)}{\pi}\;
    \langle n|
   \,\mp\Im m\,(E\pm\imath 0-H_0)^{-1}\,
    |m\rangle\,,
\qquad
b=f(E)\;,
\end{equation}
and $\Im m\,(E\pm\imath 0-H_0)^{-1}$ has a large kernel  on $\ell^2(\ZM^{d_1}\times\{0\})$ (see Proposition~\ref{prop-Greenkernel}), the set $(\psi_{m,b})_{m\in\ZM^d}$ is not a basis of $L^2(\Sigma,\nu)$ though, namely it contains many linearly dependent vectors. Next let us introduce the subspace $\Dd^\Lambda_b\subset L^2(\Sigma,\nu)$ spanned by the $\left(\psi_{m,b}\right)_{m\in\Lambda}$ (N.B. that $m$ only runs through $\Lambda$ here). Again, $\left(\psi_{m,b}\right)_{m\in\Lambda}$ is a complete set for $\Dd^\Lambda_b$, but not a basis. Furthermore, the following operators will be used:
$$
\Rr_b^\Lambda \;=\;
\sum_{m\in\Lambda} |\psi_{m,b}\rangle\langle m|
\;,
\qquad
(\Rr_b^\Lambda )^*\;=\;
\sum_{m\in\Lambda} |m\rangle\langle\psi_{m,b}|
\;.
$$ 
By \eqref{eq-scalprodL2}, or Lemma~2 and Corollary~1 of \cite{BS}, one has that
$$
(\Rr_b^\Lambda )^*\Rr_b^\Lambda  \;=\; 
   \frac{F(E)}{\pi}\;
    \Im m\,G_0^\Lambda(E-\imath 0)\,,
\qquad
b=f(E)\,.
$$
Hence
$$
\Ran (\Rr_b^\Lambda )\;=\;\Ker((\Rr_b^\Lambda )^*)^\perp\;=\;\Dd^\Lambda_b\;,
\qquad
\Ran((\Rr_b^\Lambda )^*)\;=\;\Ker(\Rr_b^\Lambda )^\perp\;=\;\Ff^\Lambda_b
\;,
$$
so that a unitary $\Pi_b^\Lambda:\Ff^\Lambda_b\to \Dd^\Lambda_b$ is given by
$$
\Pi_b^\Lambda
\;=\;
\sqrt{\frac{\pi}{F(E)}}\;
\Rr_b^\Lambda \,
    \left(
     \Im m\,G_0^\Lambda(E-\imath 0)
    \right)^{-\frac{1}{2}}\;,
\qquad
b=f(E)\;.
$$
Replacing the definition of $\Rr_b^\Lambda $ this can also be written as
\begin{equation}
\label{eq-sumproj}
\sum_{m\in\Lambda} |\psi_{m,b}\rangle\langle m|
\;=\; 
\Pi_b^\Lambda \,
   \sqrt{
    \frac{F(E)}{\pi}
         }\;
    \left(
     \Im m\,G_0^\Lambda(E-\imath 0)
    \right)^{\frac{1}{2}}\;
      P_b^\Lambda \,,
\qquad
b=f(E)\,.
\end{equation}
Furthermore, let us extend $\Pi_b^\Lambda$ to $\Pi_b^\Lambda:\ell^2(\Lambda)\to L^2(\Sigma,\nu)$ by setting $\Pi_b^\Lambda|_{(\Ff^\Lambda_b)^\perp}=0$. Then $(\Pi_b^\Lambda)^*:L^2(\Sigma,\nu)\to\ell^2(\Lambda)$ also vanishes on $(\Dd^\Lambda_b)^\perp$. Now $\Pi_b^\Lambda$ and $(\Pi_b^\Lambda)^*$ are merely partial isometries and one has $P_b^\Lambda =(\Pi_b^\Lambda)^*\Pi_b^\Lambda$ as well as $\Pi_b^\Lambda=\Pi_b^\Lambda P_b^\Lambda $, and $\Im m\,G_0^\Lambda(f^{-1}(b)\pm\imath 0)$ commutes with $P_b^\Lambda $.

\vspace{.2cm}

Finally let us introduce a partial isometry $\Pi_B^\Lambda : L^2(\RM)\otimes\ell^2(\Lambda)\to L^2(\RM)\otimes L^2(\Sigma,\nu)$ by setting
$$
\Pi^\Lambda_B\;=\;\int^\oplus \!\!db\;\Pi_b^\Lambda
\;.
$$
Then $\Ran(\Pi_B^\Lambda )=\int^\oplus \!\!db\,\Dd^\Lambda_b$ and $\Ran((\Pi_B^\Lambda )^*)=\int^\oplus \!\!db\,\Ff^\Lambda_b$. Also one has $(\Pi_B^\Lambda)^*\Pi_B^\Lambda =\int^\oplus \!\!db\,P^\Lambda_b$ and similarly $\Pi_B^\Lambda (\Pi_B^\Lambda )^*$ is equal to the direct integrals of the projections on $\Dd_b^\Lambda$. 

\vspace{.2cm}

\begin{defini}
\label{def-REI}
An operator $\widetilde{O}:L^2(\RM)\otimes L^2(\Sigma,\nu)\to L^2(\RM)\otimes L^2(\Sigma,\nu)$ in the {\rm REF} representation is called {\rm REI} representable {\rm (}w.r.t. $\Lambda$ and $H_0${\rm )} if $\widetilde{O}=\Pi_B^\Lambda (\Pi_B^\Lambda )^*\widetilde{O}=\widetilde{O}\,\Pi_B^\Lambda (\Pi_B^\Lambda )^*$, or alternatively, if $\Ran(\widetilde{O})\subset\Ran(\Pi_B^\Lambda )$ and $\Ker(\widetilde{O})\supset\Ran(\Pi_B^\Lambda )^\perp$. For any {\rm REI} representable operator $\widetilde{O}$, its {\rm REI} representation $\overset{\;\circ}{O}:L^2(\RM)\otimes\ell^2(\Lambda)\to L^2(\RM)\otimes\ell^2(\Lambda)$ is defined by 
$$
\overset{\;\circ}{O}
\;=\;
(\Pi_B^\Lambda )^*\,\widetilde{O}\,\Pi_B^\Lambda 
\;.
$$
Just as there is an {\rm EF} representation associated to the {\rm REF} representation, there is an {\rm EI} representation associated to the {\rm REI} representation.
\end{defini}

What will be of importance further below is that the REI representable operators form an algebra. Furthermore, for every REI representable operator $O$, one has
\begin{equation}
\label{eq-Ocircid}
\overset{\;\circ}{O}
\;=\;
\overset{\;\circ}{O}\,
(\Pi_B^\Lambda )^*\,\Pi_B^\Lambda 
\;=\;
(\Pi_B^\Lambda )^*\,\Pi_B^\Lambda \,
\overset{\;\circ}{O}\,
\;.
\end{equation}
Further below two different types of REI representable operators will play a role: one being operators with integral kernels in the component $L^2(\RM)$ with values in the bounded operators on $\ell^2(\Lambda)$ (this includes decaying integral kernels corresponding to compact operators in the factor $L^2(\RM)$), the other being operators having a direct integral representation in $L^2(\RM)$ with fibers given by bounded operators on $\ell^2(\Lambda)$. 

\subsection{Action of the translation group in the REF and REI representations}
\label{sec-translationaction}

The action of the translation group on $\ell^2(\ZM^d)$ is given by the unitary shifts $U_n$, $n\in\ZM^d$, defined by $U_n|m\rangle=|m-n\rangle$. Upon Fourier transform, this representation is given by multiplication operators:
$$
(\widehat{U}_n\,\psi)(k)
\;=\;
e^{\imath n\cdot k}\,\psi(k)
\;,
\qquad
\psi\in L^2(\TM^d)
\;.
$$
These operators commute with the multiplication with $\Ee$, as it should be because $H_0$ is translation invariant. Consequently their REF representation is fibered $\widetilde{U}_n=\int^\oplus db\,\widetilde{U}_{n,b}$ with unitary fibers given by
$$
(\widetilde{U}_{n,b}\,\phi)(\sigma)
\;=\;
e^{\imath n\cdot \theta_b(\sigma)}\,\phi(\sigma)
\;,
\qquad
\phi\in L^2(\Sigma,\nu)
\;.
$$
Now let $\Lambda=\ZM^{d_1}\times \{0\}$ (or, more generally, let $\Lambda$ be some subgroup of $\ZM^d$). Then $\Dd_b^\Lambda$ is invariant under the action $n_1\in \ZM^{d_1}\mapsto \widetilde{U}_{n_1,b}$ and this implies that each $\widetilde{U}_{n_1,b}$ is REI representable. Its REI representation is decomposable and particularly simple. In fact, its fibers are given by the restriction of the natural action
\begin{equation}
\label{eq-natact}
(U_{n_1}\psi)_b(m_1)\;=\;\psi_b(m_1-n_1)\;,
\qquad
\psi\in L^2(\RM)\otimes\ell^2(\Lambda)
\;,
\end{equation}
to the subspace $L^2(\RM)\otimes\Ff^\Lambda_b$.  It follows from \eqref{eq-sumproj} that this action satisfies for all $b\in\RM$ the following relation used later on
\begin{equation}
\label{eq-natact2}
\Pi^\Lambda_b \,U_{n_1}
\;=\;
\widetilde{U}_{n_1}\,\Pi^\Lambda_b
\;.
\end{equation}

\section{Deterministic results for surface scattering}
\label{sec-deterministic}

In this section, the perturbation $V$  supported by $\Lambda=\ZM^{d_1}\times\{0\}$ is fixed and therefore the index $\omega$ is suppressed on $V$ and $H$. It will be assumed throughout that $d_2=d-d_1\geq 3$ so that the bound of Corollary~\ref{coro-freeGreen} holds. 

\subsection{Basic spectral analysis of the perturbed problem}
\label{sec-spectral}

Similar as in \eqref{eq-freeresol} the perturbed resolvent is defined by $G^\Lambda(z)=\Pi\,(z-H)^{-1}\,\Pi^*$. The following formulas are well-known.

\begin{lemma}
\label{lem-Greenperturb}
For $z\in\CM\setminus\RM$,
\begin{equation}
\label{eq-Greenperturb1}
G^\Lambda(z)\;=\;
 \bigl(G_0^\Lambda(z)^{-1}-V^\Lambda\bigr)^{-1}\;=\;
  \bigl(\one-G_0^\Lambda(z)V^\Lambda\bigr)^{-1}G_0^\Lambda(z)\,,
\end{equation}
Let the $T$-matrix be defined by
\begin{equation}
\label{eq-Tmatrix}
T(z)\;=\;\Pi^*\,T^\Lambda(z)\,\Pi\,,
\hspace{2cm}
 T^\Lambda(z)\;=\;
V^\Lambda  \bigl(\one-G_0^\Lambda(z)V^\Lambda\bigr)^{-1}\;=\;
  \bigl(\one-V^\Lambda G_0^\Lambda(z)\bigr)^{-1}V^\Lambda\,.
\end{equation}
Then
\begin{equation}
\label{eq-Greenperturb2}
\frac{1}{z-H}\;=\;
 \frac{1}{z-H_0}+
  \frac{1}{z-H_0}\,T(z)\,\frac{1}{z-H_0}\,,
\end{equation}
\end{lemma}

\noindent {\bf Proof.} First of all, $G_0^\Lambda(z)^{-1}$ is invertible because, say with $\Im m(z)=\epsilon>0$,  there is a constant $C_\epsilon>0$ such that
$$
\Im m\,G^\Lambda_0(E+\imath\epsilon)
\;=\;
\Pi\,\frac{\epsilon}{(E-H_0)^2+\epsilon^2}\,\Pi^*
\;>\;C_\epsilon\,\one
\;,
$$
where it was used that $H_0$ is bounded and $E$ fixed. Now recall the general fact that for operators $A=A^*$ and $B\geq C\,\one$ on Hilbert space, the inverse of $A+\imath\,B=B^{\frac{1}{2}}(B^{-\frac{1}{2}}AB^{-\frac{1}{2}}+\imath\,\one)B^{\frac{1}{2}}$ exists and is bounded. This shows that all expressions in \eqref{eq-Greenperturb1} are well-defined, and furthermore that the inverse of $\one-G_0^\Lambda(z)V^\Lambda=G_0^\Lambda(z)(G_0^\Lambda(z)^{-1}- V^\Lambda)$ exists so that also $T^\Lambda(z)$ is well-defined. The algebraic part of the proof of all identities can now be found in Lemma~8 of \cite{BS}.
\hfill $\Box$

\vspace{.2cm}

As the subspace $\ell^2(\Lambda)=\Pi^*\ell^2(\ZM^{d_1})$ is cyclic for $H$, the spectral properties of $H$ can be read off from the boundary values of the restricted resolvent $G^\Lambda(z)$. In particular, due to \eqref{eq-Greenperturb1}, eigenvalues of $H$ must result from poles of $\bigl(\one-G_0^\Lambda(z)V^\Lambda\bigr)^{-1}$ because $G^\Lambda_0(E)$ has none by Proposition~\ref{prop-freeGreen}. Alternatively, due to \eqref{eq-Greenperturb2} eigenvalues can only result from poles of $T^\Lambda(E)$. By Corollary~\ref{coro-freeGreen}, $G^\Lambda_0(E)$ is uniformly bounded in norm and this implies the following result which can already be found in \cite{JL1}, albeit with a different proof.

\begin{proposi}
\label{prop-nopoint} Let $d_2\geq 3$. If $\|V\|< C_0$, then $H$ has no singular spectrum. 
\end{proposi}

This does not mean that there are no edge states though, but only that the edge spectrum is absolutely continuous if $\|V\|< C_0$ (see Section~\ref{sec-periodic}, for example). Let us also point out explicitly that the statement is false in dimension $d=2$ and $d_1=d_2=1$  for which it is known that there is point spectrum outside of the spectrum $\sigma(H_0)=[E_-,E_+]$ of $H_0$ \cite{JM,JL1}.

\subsection{Calculation of wave operators}
\label{sec-wavecalc}

Let $H=H_0+V$ be as described above. Then the wave operators are defined by
$$
W_\pm \;=\;
   \mbox{\rm s-}\!\!\!\lim_{t\rightarrow \pm \infty}\;
     e^{\imath Ht}\;e^{-\imath H_0t}\,.
$$
The existence of the limit can be checked by Cook's method \cite{JL1,CS}, but under the weak coupling hypothesis $\|V\|< C_0$ this also follows from the approach described now which also provides explicit formulas for the wave operators. Indeed, it follows from Proposition~11 in \cite{BS} that the Fourier transform $\widehat{W}_\pm=\Ff W_\pm\Ff^*$ of the wave operator is an integral operator of the following form
$$
\bigl((\widehat{W}_\pm-\one)\phi\bigr)(k)\;=\;
 \lim_{\epsilon\downarrow 0}\;
  \int_{\TM^d}\frac{dk'}{(2\pi)^d}\,
   \sum_{n,m\in\Lambda} \,
    \langle n|
     \,T^\Lambda(\Ee(k')\mp\imath \epsilon)\,
      |m\rangle\;
       \frac{e^{\imath(k\cdot n-k'\cdot m)}}
             {\Ee(k')\mp\imath \epsilon-\Ee(k)}\;
         \phi(k')\,.
$$
Next let us go to the REF representation, namely calculate the wave operator $\widetilde{W}_\pm=\Uu\widehat{W}_\pm\Uu^*$ which is an operator on  $L^2(\RM)\otimes L^2(\Sigma,\nu)$.  Replacing the definitions of $\Uu$ and of the states $\psi_{m,b}$, it follows that
$$
((\widetilde{W}_\pm-{\bf 1})\phi)_b \,=\,
 \lim_{\epsilon\downarrow 0}\;
  \int db'\;
   \sum_{n,m\in\Lambda} 
    |\psi_{n,b}\rangle\;
     \frac{
       \langle n|\,T^\Lambda(f^{-1}(b')\mp\imath \epsilon)\,|m\rangle
          }{f^{-1}(b')\mp\imath \epsilon-f^{-1}(b)}\;
        \langle \psi_{m,b'}|\phi_{b'}\rangle\,,
$$
where $\langle \psi_{m,b'}|\phi_{b'}\rangle$ stands for the inner product in the Hilbert space $L^2(\Sigma,\nu)$ and the integral of $b'$ carries over $\RM$. In order to shorten notations, let us write $E=f^{-1}(b)$ and $E'=f^{-1}(b')$. This can be seen as a purely formal replacement right now and does not mean that we pass from the REF to the EF representation. Thanks to \eqref{eq-sumproj}, the sums over $n$ and $m$ can be computed to give
\begin{equation}
\label{eq-Omegaformula}
((\widetilde{W}_\pm-{\bf 1})\phi)_b=
   \lim_{\epsilon\downarrow 0}
    \int \frac{db'}{\pi}\,
     \frac{F(E)^{\frac{1}{2}}\,F(E')^{\frac{1}{2}}}
          {E'\mp\imath \epsilon-E}
     \,\Pi_b^\Lambda \,
      \bigl|\Im m\,G^\Lambda_0(E)\bigr|^{\frac{1}{2}}
        \bigl(\overset{\;\circ}{O}_{\pm}\,\Pi^\Lambda_B\,\phi\bigr)_{b'},
\end{equation}
where the energy fibered operator $\overset{\;\circ}{O}_{\pm}=\int\! db\,\overset{\;\circ}{O}_{\pm,b}:L^2(\RM)\otimes \ell^2(\Lambda)\to L^2(\RM)\otimes \ell^2(\Lambda)$ with
\begin{equation}
\overset{\;\circ}{O}_{\pm,b} 
\; = \;
 \lim_{\epsilon\downarrow 0}\;
T^\Lambda(E\mp\imath \epsilon)\; 
   \bigl|\Im m\,G^\Lambda_0(E)\bigr|^{\frac{1}{2}}
\; = \;
  \Bigl(\,\one-\,V^\Lambda\,G^\Lambda_0(E\mp\imath 0)\,
\Bigr)^{-1}\;V^\Lambda\; 
     \bigl|\Im m\,G^\Lambda_0(E)\bigr|^{\frac{1}{2}}
     \,.
\label{eq-Oform}
\end{equation}
Note that there is a difference w.r.t. the definition of $\overset{\;\circ}{O}_\pm$ in \cite{BS} where a supplementary factor $(e^{\frac{b}{2}}+e^{\frac{b}{2}})^{-1}$ was introduced in order to deal with threshold singularities. Here this is not necessary because of the simplifying hypothesis $\|V\|< C_0$ which combined with Corollary~\ref{coro-freeGreen} implies that the inverse in $\overset{\;\circ}{O}_\pm$ exists and, due to Proposition~\ref{prop-freeGreen}, that 
$$
\lim_{b\to\pm\infty}
\;\overset{\;\circ}{O}_{\pm,b}
\;=\;0
\;.
$$
Now due to the formulas $\Ee(\theta_{b}(\sigma))=f^{-1}(b)=E_r+\Delta\tanh(b)$ and $F(f^{-1}(b))=\Delta\cosh^{-2}(b)$, a bit of algebra leads to
$$((\widetilde{W}_\pm-{\bf 1})\phi)_b \;=\;
   \Pi_b^\Lambda \;
    \bigl|\Im m\;G^\Lambda_0(f^{-1}(b))\bigr|^{\frac{1}{2}}\,
     \int \frac{db'}{\pi}\;
      \frac{1}{\sinh(b'-b)\mp\imath 0} \;
        (\overset{\;\circ}{O}_{\pm}\,(\Pi^\Lambda_B)^*\,\phi)_{b'}\,.
$$
\noindent In the previous formula, $\overset{\;\circ}{O}_{\pm}(\Pi^\Lambda_B)^*\phi$ is a vector in  the Hilbert space $L^2(\RM)\otimes \ell^2(\Lambda)$. As previously let ${A}=-\imath\partial_b$ be the generator of the translation group in $L^2(\RM)\otimes L^2(\Sigma,\nu)$ as well as $L^2(\RM)\otimes \ell^2(\Lambda)$. Changing the integration variable $b'$ to $u=b'-b$ leads to  $(\overset{\;\circ}{O}_{\pm}(\Pi^\Lambda_B)^*\phi)_{u+b}=(e^{\imath {A}u}\overset{\;\circ}{O}_{\pm}(\Pi^\Lambda_B)^*\phi)_{b}$. Hence
$$
\Big((\widetilde{W}_\pm-{\bf 1})\phi\Big)_b=\,
  \Pi_b^\Lambda \,
   \bigl|\Im m\;G^\Lambda_0(f^{-1}(b))\bigr|^{\frac{1}{2}}\!
     \int\frac{du}{\pi}\,
       \frac{1}{\sinh(u)\mp\imath 0} \,
         \left(e^{\imath {A}u}\,
\overset{\;\circ}{O}_{\pm}\,(\Pi^\Lambda_B)^*\,\phi
         \right)_{b}\!.
$$
\noindent Now the identity
\begin{equation}
\label{eq-integralid}
\int\frac{du}{\imath\pi}\;
 \frac{1}{\sinh\bigl(u\bigr)\mp\imath 0}\;
    e^{\imath {A}u}\;=\;
     \pm\,\one\;+\;
\tanh\bigl(\tfrac{\pi}{2}{A}\,\bigr)
\;,
\end{equation}
implies the following result:

\begin{theo}
\label{theo-waveop}
Let $d\geq 3$ and $\|V\|<C_0$. Then the {\rm REF} representation of the wave operators is
$$
\widetilde{W}_\pm -\one\,=\, 
   \imath\,\Pi_{{B}}^\Lambda\,
    \bigl|\Im m\,G^\Lambda_0(f^{-1}({{B}}))\bigr|^{\frac{1}{2}}\,
     \left(
     \pm\one\,+\,
\tanh\bigl(\tfrac{\pi}{2}{A}\,\bigr)
     \right)\,
T^\Lambda(f^{-1}({B})\mp\imath 0)\, 
     \bigl|\Im m\,G^\Lambda_0(f^{-1}({B}))\bigr|^{\frac{1}{2}}
      \,(\Pi_{{B}}^\Lambda)^*\,.
$$
In particular, the difference of wave operator and identity is {\rm REI} representable.
\end{theo}

Theorem~\ref{theo-LevinsonGen} will be deduced from this formula. Before going on, let us briefly comment on which technical difficulties have to be overcome in order to extend the formula to the strong coupling regime. In such a situation the existence of the inverse in \eqref{eq-Oform} has to follow from other reasons. First of all, let us set
$$
\alpha_b\;=\;\bigl(\pi^VV^\Lambda(\pi^V)^*\bigr)^{-1}-\pi^V\Re e\,G^\Lambda_0(E)\,(\pi^V)^*
\qquad
\beta_b\;=\;\bigl|\Im m\,G^\Lambda_0(E)\bigr|^{\frac{1}{2}}\,(\pi^V)^*
\;,
$$ 
where as usual $b=f(E)$ and the inverse in $\alpha_b$ is supposed to exist. Then $\alpha_b$ is a self-adjoint operator on $\Ran(V^\Lambda)$ and $\beta_b:\Ran(V^\Lambda)\to\ell^2(\Lambda)$, and one has
$$
\overset{\;\circ}{O}_{\pm,b}
\;=\;
\Bigl(\,\alpha_b\mp\imath\,\beta_b^*\beta_b\,
\Bigr)^{-1}\;\beta^*_b
\;.
$$
Next one has $\Ker(\alpha_b\mp\imath\,\beta_b^*\beta_b)=\Ker(\alpha_b)\cap\Ker(\beta_b)$ (see {\it e.g.} Appendix~B of \cite{BS}). But $\Ran(\beta_b^*)=\Ker(\beta_b)^\perp$ so that $\overset{\;\circ}{O}_{\pm,b}$ is well-defined. Now proving that it is bounded appears to be difficult (the techniques of Appendix~B of \cite{BS} only apply to finite $\Lambda$), and a uniform bound (in $b$) can only be obtained for $\cosh(b)^{-1}\overset{\;\circ}{O}_{\pm,b}$ under supplementary hypothesis on the nature of the half-bound states (see \cite{BS}). Then factor $\cosh(b)^{-1}$ has to and actually can be compensated (see again \cite{BS}). Let us point out that $\one+2\imath\beta_b\overset{\;\circ}{O}_{\pm,b}$ is unitary which implies that $\beta_b\overset{\;\circ}{O}_{\pm,b}$ is uniformly bounded, but a factor $(\beta_b)^{-1}$ cannot be compensated in the expression for the wave operators. All these issues may not only be of technical nature, but are possibly also connected to half-bound state corrections to \eqref{eq-main}.

\subsection{Scattering and time delay operator}
\label{sec-Smatrix}

Applying the invariance principle $S=\mbox{s-}\lim_{t\to\infty} e^{\imath tB}W_-e^{-\imath tB}$ to the formula in Theorem~\ref{theo-waveop} now implies the following formula for the $S$-matrices. 

\begin{theo}
\label{theo-scatop}
Let $d\geq 3$ and $\|V\|< C_0$. The scattering operator in the {\rm REF} representation is fibered $\widetilde{S}=\int^\oplus \!\!db\,\widetilde{S}_b$ and {\rm REI} representable with unitary fibers $\widetilde{S}_b=\Pi_{b}^\Lambda \overset{\;\circ}{S}_b (\Pi_{b}^\Lambda)^*$ given by
$$
\overset{\;\circ}{S}_b \,=\, 
\one-\,2\,   \imath\,
    \bigl|\Im m\,G^\Lambda_0(E)\bigr|^{\frac{1}{2}}\,
    \Bigl(\,\one-\,V^\Lambda\,G^\Lambda_0(E+\imath 0)\,
\Bigr)^{-1}\;V^\Lambda\;
     \bigl|\Im m\,G^\Lambda_0(E)\bigr|^{\frac{1}{2}}\,,
\qquad
b=f(E)\;.
$$
Furthermore
\begin{equation}
\label{eq-Slimit}
\lim_{|b|\to\infty} \,\overset{\;\circ}{S}_b\;=\;
\one
\;.
\end{equation}
\end{theo}

The assymptotics \eqref{eq-Slimit} follow from the stated fromula when Proposition~\ref{prop-freeGreen} and \eqref{eq-GreenGlobFourier} are taken into account. The time delay operator is by definition $T=-\imath \,S^*[A,S]$ (this is also denoted by $T$, just as the $T$-matrix, but hopefully no confusion results from this). In the REF representation is given by
$$
\widetilde{T}=\int^\oplus \!\!db\;\widetilde{T}_b\;,
\qquad
\widetilde{T}_b\;=\;\frac{1}{\imath}\;\widetilde{S}_b^*\,\partial_b\,\widetilde{S}_b
\;.
$$
Using \eqref{eq-Ocircid} and the unitarity of $\overset{\;\circ}{S}_b$, one finds
$$
\widetilde{T}_b
\;=\;
\frac{1}{\imath}\;
\Pi_{b}^\Lambda \,(\overset{\;\circ}{S}_b)^* \,(\Pi_{b}^\Lambda)^*
\,\partial_b \Pi_{b}^\Lambda \,\overset{\;\circ}{S}_b\, (\Pi_{b}^\Lambda)^*
+
\frac{1}{\imath}\;
\Pi_{b}^\Lambda \,(\overset{\;\circ}{S}_b)^*
\,\partial_b \overset{\;\circ}{S}_b\, (\Pi_{b}^\Lambda)^*
+
\frac{1}{\imath}\;
\Pi_{b}^\Lambda \,\partial_b (\Pi_{b}^\Lambda)^*
\;.
$$
Thus
\begin{equation}
\label{eq-TDtrace}
\Tr_{L^2(\Sigma,\nu)}(\widetilde{T}_b)
\;=\;
\frac{1}{\imath}\;
\Tr_{\ell^2(\Lambda)}\bigl(
(\overset{\;\circ}{S}_b)^*
\,\partial_b \overset{\;\circ}{S}_b\bigr)
\;.
\end{equation}

\subsection{Projection on the surface states}
\label{sec-projsurface}

Let us begin by recalling an important structural result from \cite{JL1,JL2} which actually defines the projection on the surface states.

\begin{theo}
\label{theo-JL} {\rm \cite{JL1,JL2}} 
Let $H=H_0+V$ be as described in {\rm Section~\ref{sec-intro}}. Then the wave operators have common range $\Ran(W_+)=\Ran(W_-)$. Moreover, $P_{\mbox{\tiny\rm bulk}}=W_\pm W_\pm^*$ is an orthogonal projection on $\Ran(W_\pm)$ characterized by
$$
\Ran(P_{\mbox{\tiny\rm bulk}})
\;=\;
\left\{
\psi\in\ell^2(\ZM^{d})\;\left|\;\int^\infty_0 dt\;\|\Pi_1\,e^{-\imath t H}\,\psi\|^2\,<\,\infty\right.
\right\}^{\mbox{\rm\tiny cl}}
\;,
$$
where $\mbox{\rm cl}$ denotes the closure and $\Pi_1$ is a the projection on a strip of size $1$ around $\Lambda$. Then $P_{\mbox{\tiny\rm ss}}=\one-P_{\mbox{\tiny\rm bulk}}=\one-W_\pm W_\pm^*$ is called the projection on the surface states.
\end{theo}

\noindent {\bf Remark} In \cite{JL2} it is actually shown that
$$
\Ran(P_{\mbox{\tiny\rm bulk}})
\;=\;
\left\{
\psi\in\ell^2(\ZM^{d})\;\left|\;\int^\infty_0 dt\;\|\Pi_R\,e^{-\imath t H}\,\psi\|^2\,<\,\infty\right.
\;\;\forall \;\,R\in\NM
\right\}^{\mbox{\rm\tiny cl}}
\;,
$$
where $\Pi_R$ is the projection on a strip of with $R$ around $\Lambda$. However, the proof also gives the above result.
\hfill $\diamond$

\vspace{.2cm}

The projection on the surface states is given by $P_{\mbox{\tiny\rm ss}}=\one-W_\pm W_\pm^*$ which can be rewritten as
\begin{equation}
\label{eq-PSurdecomp}
P_{\mbox{\tiny\rm ss}}
\;=\;
(\one-W_\pm )\,+\,(\one-W_\pm )^*
\,-\,(\one-W_\pm )(\one-W_\pm )^*
\;.
\end{equation}
If follows from Theorem~\ref{theo-waveop} that $P_{\mbox{\tiny\rm ss}}$ is REI representable. Using the formula in Theorem~\ref{theo-waveop} one can now also write out a somewhat lengthy explicit formula for the REI representation of $P_{\mbox{\tiny\rm ss}}$. This allows to study the boost in $A$ and $B$, namely the vanishing of the imaginary part of the free resolvent at the band edges implies
$$
\lim_{|t|\to\infty}\;e^{\imath At}\,\overset{\;\circ}{P}_{\mbox{\rm\tiny ss}}\,e^{-\imath At}\
\;=\;
0\;,
$$
and because $\pm\one+\tanh\bigl(\tfrac{\pi}{2}{A}\,\bigr)$ vanishes at $\mp\infty$, choosing the corresponding sign in \eqref{eq-PSurdecomp} leads to
$$
\lim_{|t|\to\infty}\;e^{\imath Bt}\,\overset{\;\circ}{P}_{\mbox{\rm\tiny ss}}\,e^{-\imath Bt}
\;=\;
0\;.
$$
Instead of using both formulas in \eqref{eq-PSurdecomp} one can also verify this directly on one of the formulas. For example, if one uses $W_+$, then the limit $t\to\infty$ is given by
$$
\lim_{t\to\infty}\;e^{\imath Bt}\,\overset{\;\circ}{P}_{\mbox{\rm\tiny ss}}\,e^{-\imath Bt}
\;=\;
2\,
    \bigl|\Im m\,G^\Lambda_0\bigr|^{\frac{1}{2}}\,
     \left(
2\,
T^\Lambda\, \bigl|\Im m\,G^\Lambda_0\bigr|\,(T^\Lambda)^*
+\imath\, T^\Lambda-\imath\, (T^\Lambda)^*
\right)\,
 \bigl|\Im m\,G^\Lambda_0\bigr|^{\frac{1}{2}}
\;=\;0
\;,
$$
where the argument $f^{-1}({{B}})$ in $G^\Lambda_0$ and $T^\Lambda(\,.\,-\imath 0)$ was dropped, and the second equality follows from \eqref{eq-Tmatrix} after a short calculation. From these asymptotics one concludes that $\overset{\;\circ}{P}_{\mbox{\rm\tiny ss}}$ is compact in the rescaled energy variable. As all compact projections are traceclass this implies the following result showing that the partial trace of the REF and REI representations of ${P}_{\mbox{\rm\tiny ss}}$ over the fiber $L^2(\RM)$ are well-defined operators on $L^2(\Sigma,\nu)$ and $\ell^2(\Lambda)$ respectively. 

\begin{proposi}
\label{prop-Psurtraceclass}
The {\rm REF} and {\rm REI} representations of the projection ${P}_{\mbox{\rm\tiny ss}}$ on the surface states is traceclass in the fiber $L^2(\RM)$ corresponding to the rescaled energy variable.
\end{proposi}

\section{Levinson's theorem in the case of finite support $\Lambda$}
\label{sec-finitesupport}

If $\Lambda$ is finite and $d\geq 3$, all arguments of Section~\ref{sec-deterministic} leading to the formulas for the wave operator (Theorem~\ref{theo-waveop}) and the scattering operator (Theorem~\ref{theo-scatop}) carry over if Corollary~\ref{coro-freeGreen} is replaced by Proposition~\ref{prop-freeGreen}. In \cite{BS} both formulas were even proved without the assumption on the weakness of the perturbation (but a technical assumption on the nature of the threshold resonances). Then there may be bound states as well as embedded eigenvalues and half-bound states.  Based on these analytical results, the Levinson theorem was deduced. As a preparation to the surface scattering problem in Section~\ref{sec-algebras} and in order to advertise the advantages of the REI representation, let us prove Levinson's theorem for a perturbation of finite support $\Lambda$, focussing on the situation without embedded eigenvalues and half-bound states. In particular, the $S$-matrix then converges to the identity as the energy converges to the band edges. Furthermore the projection $P=W_\pm W_\pm^*-\one$ is on the eigenspace of all eigenvalues (bound states) of $H$. Following the idea of \cite{KR}, the Levinson theorem is obtained as an index theorem of an adequate exact sequence of C$^*$-algebras. This sequence was already used in \cite{GI} for a different purpose. The algebras contain the REI representation of the operators of scattering theory and are smaller than the algebras used in \cite{BS}. Let $|\Lambda|=L$ and denote by $\mbox{\rm Mat}(L,\CM)$ the complex $L\times L$ matrices, which are all the bounded operators on $\ell^2(\Lambda)$. Let $\Jj$ be the C$^*$-algebra generated by operators of the form $f({A})\otimes M$ and $g({B})\otimes M'$ with $f,g\in C_0(\RM)$, operators $M,M'\in\mbox{\rm Mat}(L,\CM)$. If $\Kk=C_0(A,B)$ denotes the compact operators on $L^2(\RM)$, then $\Jj=\Kk\otimes \mbox{\rm Mat}(L,\CM)$. Let $\Ee=C_\infty(A,B)$ denote the extension of $\Jj$ obtained by allowing $f$ and $g$ to have nonzero finite limits at $\pm\infty$. Evaluation at infinity of $\Ee$ gives the algebra $\Aa$ which is the subalgebra of $\bigl(C_\infty({A})\oplus C_\infty({B})\oplus C_\infty({A})\oplus C_\infty({B})\bigr)\otimes \mbox{\rm Mat}(L,\CM)$ of fibered operators having coinciding limits in the four corners.  Thus one obtains the following short exact sequence  of C$^*$-algebras
\begin{equation}
\label{eq-exactseq}
0\;\to\;\Jj\;
    \hookrightarrow\;\Ee\;
     \overset{\mbox{\rm\tiny ev}}{\to}\;\Aa\;\to\;0\;.
\end{equation}
Now it follows from the results above that the REI representations of the projection $P$ on the bound, the wave operators and the scattering operator are respectively in $\Jj$, $\Ee$ and $\Aa$ respectively. More precisely, the REI representation of $\one\otimes \overset{\;\circ}{S}\otimes\one\otimes\one$ lies in $\Ee$. Furthermore, by the invariance principle $\overset{\;\circ}{W}_-$ is its lift and thus its image under the $K$-theoretic index map is the class of $\overset{\;\circ}{W}_-(\overset{\;\circ}{W}_-)^*-(\overset{\;\circ}{W}_-)^*\overset{\;\circ}{W}_-=\overset{\;\circ}{P}$ (all in REI representation), similar as in \cite{KR,BS}. As $\overset{\;\circ}{P}$ is a compact projection, it is finite dimensional and of dimension
$$
\Tr_{L^2(\RM)\otimes \CM^L}(\overset{\;\circ}{P})
\;=\;
\int db\;\Tr_L(\overset{\;\circ}{P}_{b,b})
\;=\;
\int db\;\Tr_{L^2(\Sigma,\nu)}(\widetilde{P}_{b,b})
\;=\;
\Tr_{L^2(\RM)\otimes L^2(\Sigma,\nu)}(\widetilde{P})
\;=\;
\Tr_{\ell^2(\ZM^d)}(P)
\;.
$$
Furthermore, this dimension is (up to a sign) equal to the Fredholm index of $\one\otimes \overset{\;\circ}{S}\otimes\one\otimes\one$ which is equal to that of $\overset{\;\circ}{S}$. By a Gohberg-Krein type theorem 
$$
\mbox{\rm Ind}(\overset{\;\circ}{S})
\,=\,
\frac{1}{2\pi\imath}\,
\int \!db\,\Tr_{\ell^2(\Lambda)}
\left((\overset{\;\circ}{S}_b)^*\partial_b \overset{\;\circ}{S}_b\right)
\,=\,
\frac{1}{2\pi\imath}\,
\int^{E_+}_{E_-} \!dE\,\Tr_{\ell^2(\Lambda)}
\left((\overset{\;\circ}{S}_E)^*\partial_E \overset{\;\circ}{S}_E\right)
$$
where the second equality follows from the change of variables $b=f(E)$. 
Combining these equalities gives Levinson's theorem:

\begin{theo}
\label{theo-Levfinite}
Suppose that $\Lambda$ is finite and that there are no half-bound states and no embedded eigenvalues. Then the number of bound states is equal to
$$
\Tr_{\ell^2(\ZM^d)}(P)
\;=\;
-\,\frac{1}{2\pi\imath}\,
\int^{E_+}_{E_-} \!dE\,\Tr_{\ell^2(\Lambda)}
\left((\overset{\;\circ}{S}_E)^*\partial_E \overset{\;\circ}{S}_E\right)
\;.
$$
\end{theo}

Using the results of \cite{BS} it is also possible to include corrections resulting from half-bound states and to deal with embedded eigenvalues. Even without this generalization, Levinson's theorem is stated somewhat differently in \cite{BS}. For once, there was a mistake in the last line of the proof, but disregarding this step the formula in \cite{BS} is
$$
\Tr_{\ell^2(\ZM^d)}(P)
\;=\;
-\,\frac{1}{2\pi\imath}\,
\int^{E_+}_{E_-} \!dE\,\Tr_{L^2(\Sigma,\nu)}
\left((\widetilde{S}_E)^*\partial_E \widetilde{S}_E\right)
\;.
$$
This indeed coincides with Theorem~\ref{theo-Levfinite} due to the identity \eqref{eq-TDtrace}. The proof of Theorem~\ref{theo-LevinsonGen} proceeds exactly along the same lines, except that $\mbox{\rm Mat}(L,\CM)$ is replaced by the algebra of covariant operators on $\Lambda=\ZM^{d_1}\times\{0\}$ and the trace by the trace per unit volume.

\section{Random surface perturbations}
\label{sec-algebras}

This section considers a covariant family $(V_\omega)_{\omega\in\Omega}$ of surface perturbations supported by $\Lambda=\ZM^{d_1}\times\{0\}$. Then the associated scattering theory also has covariance properties and this allows to construct adequate operator C$^*$-algebras which are the crucial ingredients of the proof of Theorem~\ref{theo-LevinsonGen}. In this section, all objects carry the index $\omega$ to indicate the dependence on the surface perturbation.

\subsection{Covariance properties of surface scattering}
\label{sec-covariance}

Let $(\Omega,T,\ZM^{d_1},\PM)$ be a compact dynamical system with invariant and ergodic probability measure $\PM$.  By definition, a family $(O_\omega)_{\omega\in\Omega}$ of bounded operators $O_\omega$ on $\ell^2(\ZM^{d_1})$ or $\ell^2(\ZM^{d})$ is called  covariant w.r.t. the shift on $\ZM^{d_1}$ or along the hyperplane $\Lambda=\ZM^{d_1}\times\{0\}$ if and only if
\begin{equation}
\label{eq-covariance}
O_{T^{n_1}\omega}\;=\;U_{n_1}\,O_\omega\, U_{n_1}^*
\;,
\qquad
n_1\in\ZM^{d_1}\;,
\end{equation}
where $U_{n_1}$ is the translation in $\ell^2(\ZM^{d_1})$ or $\ell^2(\ZM^{d})$ by $n_1\in\ZM^{d_1}$. By hypothesis, the family $(V_\omega)_{\omega\in\Omega}$ of surface perturbations is a $\ZM^{d_1}$-covariant family on $\ell^2(\ZM^d)$ in this sense, and the family of restrictions $(V^\Lambda_\omega)_{\omega\in\Omega}$ on $\ell^2(\ZM^{d_1})$. It follows that the Hamiltonians $(H_\omega)_{\omega\in\Omega}$ with $H_0+V_\omega$, their resolvents as well as the restrictions of the resolvents $G^\Lambda_\omega(z)$ are also covariant families. Furthermore, the wave operators, scattering operator and surface projection are $\ZM^{d_1}$-covariant. This can be checked either directly from the definitions, or alternatively from the formulas deduced in Section~\ref{sec-deterministic} and the fact that the $T$-matrix $T^\Lambda_\omega(z)$ defined in \eqref{eq-Tmatrix} is also covariant.

\vspace{.2cm}

Now let us suppose that a $\ZM^{d_1}$-covariant family $(O_\omega)_{\omega\in\Omega}$ of operators on $\ell^2(\ZM^d)$ is REI representable on $\Lambda$.  Then their REI representation $\overset{\;\circ}{O}=(\overset{\;\circ}{O}_\omega)_{\omega\in\Omega}$ satisfies \eqref{eq-covariance}, wtih unitaries $U_{n_1}$ are given by \eqref{eq-natact}. All this is given for the wave, scattering and time delay operator (which are energy fibered) and the projection on the surface states (which is an integral operator in the energy variable).

\vspace{.2cm}

As usual \cite{Bel}, the covariant operators form a crossed product C$^*$-algebra $C(\Omega)\rtimes \ZM^{d_1}$, or, more precisely, are given by representations of this algebra. Elements of this algebra can be approximated by compactly supported functions $O(\omega,n)$ on $\Omega\times\ZM^{d_1}$, which provide covariant operators by the identity $O(\omega,n)=\langle 0|O_\omega|n\rangle$. The reader is referred to \cite{Bel} for a detailed description of the formalism. All that is needed here is that there is a normalized trace $\Tt_1$ on $C(\Omega)\rtimes \ZM^{d_1}$ defined by
$$
\Tt_1(O)\;=\;\EE_\PM\;O(\omega,0)\;=\;\EE_\PM\;\langle 0|O_\omega|0\rangle
\;,
\qquad
O\in C(\Omega)\rtimes \ZM^{d_1}\;.
$$
By Birkhoff's ergodic theorem, this is $\PM$-almost surely equal to the trace per unit volume of the $O_\omega$:
$$
\Tt_1(O)
\;=\;
\lim_{N\to\infty}
\;\frac{1}{(2N+1)^{d_1}}\,\Tr_{\ell^2(\ZM^{d_1})}(\chi_N\,O_\omega)
\;,
$$
where $\chi_N=\sum_{|n|\leq N}|n\rangle\langle n|$ is the projection on the square $[-N,N]^{d_1}$.

\subsection{Exact sequence for surface scattering}
\label{sec-short}

The aim of this section is to construct a short exact sequence $0\to\Jj\hookrightarrow\Ee\overset{\mbox{\rm\tiny ev}}{\to}\Aa\to 0$ of C$^*$-algebras such that the REI representation of the covariant family $\Psur=(P_{\mbox{\rm\tiny ss},\omega})_{\omega\in\Omega}$ of surface projections is in $\Jj$, the REI representation of  the scattering operator $S=(S_{\omega})_{\omega\in\Omega}$ is part of the algebra $\Aa$ and the REI representation of  the wave operators $W_\pm=(W_{\pm,\omega})_{\omega\in\Omega}$ are in $\Ee$. This is achieved in a manner completely analogous to the case of a finite $\Lambda$ described in Section~\ref{sec-finitesupport}, except that the algebra of all operators on $\ell^2(\Lambda)$ is replaced by the algebra of covariant operators given by the crossed product $C(\Omega)\rtimes \ZM^{d_1}$. Hence the exact sequence is given by the C$^*$-algebras $\Jj=C_0(A,B)\otimes C(\Omega)\rtimes \ZM^{d_1}$ and $\Ee=C_\infty(A,B)\otimes C(\Omega)\rtimes \ZM^{d_1}$ as well as the algebra $\Aa$ which is the subalgebra of $\bigl(C_\infty({A})\oplus C_\infty({B})\oplus C_\infty({A})\oplus C_\infty({B})\bigr)\otimes C(\Omega)\rtimes \ZM^{d_1}$ of fibered operators having coinciding limits in the four corners. This exact sequence satisfies the requirements above and one has, by the same arguments as in \cite{KR,BS} and Section~\ref{sec-finitesupport}, in the sense of $K$-theory associated to the above exact sequence of C$^*$-algebras
$$
\mbox{\rm Ind}([\overset{\;\circ}{S}]_1)
\;=\;
[\one-(\overset{\;\circ}{W}_-)^*\overset{\;\circ}{W}_-]_0
-
[\one-\overset{\;\circ}{W}_-(\overset{\;\circ}{W}_-)^*]_0
\;=\;
-\,[\overset{\;\circ}{P}_{\mbox{\rm\tiny ss}}]_0
\;.
$$
Evaluating this (using the Gohberg-Krein theorem on $L^2(\RM)$ as in \cite{KR2} tensorized with an algebra equipped with the everywhere defined trace $\Tt_1$) gives
\begin{equation}
\label{eq-firststep}
\Tt_1\,\Tr_{L^2(\RM)}\bigl(\overset{\;\circ}{P}_{\mbox{\rm\tiny ss}}\bigr)
\;=\;
-\,
\frac{1}{2\pi\imath}\int^\infty_{-\infty} db\;
\Tt_1\bigl((\overset{\;\circ}{S}_b)^*\,\partial_b\,\overset{\;\circ}{S}_b\bigr)
\;=\;
-\,\frac{1}{2\pi\imath}\int^{E_+}_{E_-} dE\;
\Tt_1\bigl((\overset{\;\circ}{S}_E)^*\,\partial_E\,\overset{\;\circ}{S}_E\bigr)
\;,
\end{equation}
where $\Tr_{L^2(\RM)}$ the trace on the compact operators $C_0(A,B)$ represented on $L^2(\RM)$ (which exists due to Proposition~\ref{prop-Psurtraceclass}) and $\Tt_1$ is the trace per unit volume on $C(\Omega)\rtimes \ZM^{d_1}$. The equality \eqref{eq-firststep} is the main step in the proof of Theorem~\ref{theo-LevinsonGen}. It remains to show that the l.h.s. of \eqref{eq-firststep} is equal to the l.h.s. of the equation in Theorem~\ref{theo-LevinsonGen}. This is the object of the next section which then concludes the proof of Theorem~\ref{theo-LevinsonGen}.

\subsection{Traces of covariant REI representable operators}
\label{sec-traces}

\begin{proposi}
\label{prop-traceconnect}
Let $O=(O_\omega)_{\omega\in\Omega}$ be a $\ZM^{d_1}$-covariant family of {\rm REI} representable positive operators on $\ell^2(\ZM^d)$. Then, if one of the two sides is finite,
$$
\Tt_1\,\Tr_2(O)
\;=\;
\Tt_1\,\Tr_{L^2(\RM)}\bigl(\overset{\;\circ}{O}\bigr)
\;.
$$
\end{proposi}

\noindent {\bf Proof.} In the calculation below, sums over $n_1,n'_1,m_1$ are over $\ZM^{d_1}$, while those over $n_2$ over $\ZM^{d_2}$. Limits are freely exchanged which is possible due to positivity. From the definition,
$$
\Tt_1\,\Tr_2(O)
\;=\;
\EE_\PM\;\sum_{n_2}\;\langle 0,n_2|O_\omega|0,n_2\rangle
\;=\;
\EE_\PM\;
\Tr_{\ell^2(\ZM^d)}
\left(O_\omega\;
\sum_{n_2}\;|0,n_2\rangle\langle 0,n_2|
\right)
\;.
$$
Passing consecutively to the REF and REI representations, one has
\begin{eqnarray*}
\Tt_1\,\Tr_2(O)
& = &
\EE_\PM\;
\Tr_{L^2(\RM)\otimes L^2(\Sigma,\nu)}
\left(\widetilde{O}_\omega\;
\sum_{n_2}\;|\psi_{(0,n_2)}\rangle\langle \psi_{(0,n_2)}|
\right)
\\
& = &
\EE_\PM\;\int db\int db'\;
\Tr_{L^2(\Sigma,\nu)}
\left(\widetilde{O}_{\omega,b,b'}\;
\sum_{n_2}\;|\psi_{(0,n_2),b'}\rangle\langle \psi_{(0,n_2),b}|
\right)
\\
& = &
\EE_\PM\;\int db\int db'\;
\Tr_{\ell^2(\Lambda)}
\left(\overset{\;\circ}{O}_{\omega,b,b'}\;
\sum_{n_2}\;(\Pi^\Lambda_{b'})^*\,|\psi_{(0,n_2),b'}\rangle\langle \psi_{(0,n_2),b}|\,\Pi^\Lambda_{b}
\right)
\;.
\end{eqnarray*}
Next let us write out the trace and product in $\ell^2(\Lambda)$ explicitly:
$$
\Tt_1\,\Tr_2(O)
\;=\;
\EE_\PM\;\int db\int db'\;
\sum_{n_1,n'_1}\,
\langle n_1|\,\overset{\;\circ}{O}_{\omega,b,b'}\,|n'_1\rangle\;
\sum_{n_2}\;
\langle n'_1|\,(\Pi^\Lambda_{b'})^*\,|\psi_{(0,n_2),b'}\rangle\langle \psi_{(0,n_2),b}|\,\Pi^\Lambda_{b}\,|n_1\rangle
\;.
$$
Exchanging limits and using the covariance relation \eqref{eq-covariance} as well as the invariance of $\PM$,
$$
\Tt_1\,\Tr_2(O)
\;=\;
\int db\int db'\;
\sum_{n_1,m_1}\,
\EE_\PM\;
\langle 0|\,\overset{\;\circ}{O}_{\omega,b,b'}\,|m_1\rangle\;
\sum_{n_2}\;
\langle n_1+m_1|\,(\Pi^\Lambda_{b'})^*\,|\psi_{(0,n_2),b'}\rangle\langle \psi_{(0,n_2),b}|\,\Pi^\Lambda_{b}\,|n_1\rangle
\;.
$$
where $m_1=n_1'-n_1$. Due to \eqref{eq-natact2}, one now has
$$
\Tt_1\,\Tr_2(O)
\;=\;
\int db\int db'\;
\sum_{m_1}\,
\EE_\PM\;
\langle 0|\,\overset{\;\circ}{O}_{\omega,b,b'}\,|m_1\rangle\;
\sum_{n_1,n_2}\;
\langle m_1|\,(\Pi^\Lambda_{b'})^*\,|\psi_{(n_1,n_2),b'}\rangle\langle \psi_{(n_1,n_2),b}|\,\Pi^\Lambda_{b}\,|0\rangle
\;,
$$
so that by \eqref{eq-deltarel}
$$
\Tt_1\,\Tr_2(O)
\;=\;
\int db\;
\sum_{m_1}\,
\EE_\PM\;
\langle 0|\,\overset{\;\circ}{O}_{\omega,b,b}\,|m_1\rangle\;
\langle m_1|\,(\Pi^\Lambda_{b})^*\,\Pi^\Lambda_{b}\,|0\rangle
\;.
$$
By the REI representability, $\overset{\;\circ}{O}_{\omega,b,b}(\Pi^\Lambda_{b})^*\,\Pi^\Lambda_{b}=\overset{\;\circ}{O}_{\omega,b,b}$ so that the claim follows.
\hfill $\Box$


\end{document}